\documentclass[conference]{IEEEtran}
\IEEEoverridecommandlockouts
\usepackage{cite}
\usepackage{amsmath,amssymb,amsfonts}
\usepackage{footnote}
\usepackage{algorithmic}
\usepackage{graphicx}
\usepackage{textcomp}
\usepackage{xcolor}
\usepackage{lipsum}
\def\BibTeX{{\rm B\kern-.05em{\sc i\kern-.025em b}\kern-.08em
    T\kern-.1667em\lower.7ex\hbox{E}\kern-.125emX}}
\begin{document}
\newcommand{\changed}[1]{{\color{blue}#1}}

\title{Performance Analysis of Downlink MIMO-NOMA Systems Over Weibull Fading Channels\\
\thanks{
The work of L. P. J. Jiménez was supported by Eldorado Research Institute.\\
The work of F.~D.~A.~García was supported by the São Paulo Research Foundation (FAPESP) under Grant 2021/03923-9.}
}

\makeatletter
\newcommand{\linebreakand}{%
  \end{@IEEEauthorhalign}
  \hfill\mbox{}\par
  \mbox{}\hfill\begin{@IEEEauthorhalign}
}
\makeatother

\author{
\IEEEauthorblockN{Lenin Patricio Jiménez Jiménez}
\IEEEauthorblockA{\textit{Dept. of Communications} \\
\textit{University of Campinas}\\
Campinas, Brazil \\
l264366@dac.unicamp.br}
\and
\IEEEauthorblockN{Fernando Darío Almeida García}
\IEEEauthorblockA{\textit{Dept. of Communications} \\
\textit{University of Campinas}\\
Campinas, Brazil \\
ferdaral@decom.fee.unicamp.br}
\and
\IEEEauthorblockN{Maria Cecilia Luna Alvarado}
\IEEEauthorblockA{\textit{Dept. of Communications} \\
\textit{University of Campinas}\\
Campinas, Brazil\\
m264371@dac.unicamp.br}
\linebreakand
\IEEEauthorblockN{Gustavo Fraidenraich}
\IEEEauthorblockA{\textit{Dept. of Communications} \\
\textit{University of Campinas}\\
Campinas, Brazil\\
gf@decom.fee.unicamp.br}
\and
\IEEEauthorblockN{Michel Daoud Yacoub}
\IEEEauthorblockA{\textit{Dept. of Communications} \\
\textit{University of Campinas}\\
Campinas, Brazil \\
mdyacoub@unicamp.br}
\and
\IEEEauthorblockN{José Cândido S. Santos Filho}
\IEEEauthorblockA{\textit{Dept. of Communications} \\
\textit{University of Campinas}\\
Campinas, Brazil \\
candido@decom.fee.unicamp.br}
\and
\IEEEauthorblockN{Eduardo Rodrigues de Lima}
\IEEEauthorblockA{\textit{Hardware Dept.} \\
\textit{Eldorado Research Institute}\\
Campinas, Brazil \\
eduardo.lima@eldorado.org.br}
}

\maketitle

\begin{abstract}
This work analyzes the performance of a downlink multi-user multiple-input multiple-output (MU-MIMO) non-orthogonal multiple access (NOMA) communications system. To reduce hardware complexity and exploit antenna diversity, we consider a transmit antenna selection (TAS) scheme and equal-gain combining (EGC) receivers. Further, we consider Weibull-distributed fading channels to account for non-linearities of the propagation medium and to cover, as special cases, important fading scenarios such as Rayleigh and exponential models. Performance metrics such as the outage probability (OP) and the average bit error rate (ABER) are derived in an exact manner. An asymptotic analysis for the OP and for the ABER is also carried out. Moreover, we obtain exact expressions for the probability density function (PDF) and the cumulative distribution function (CDF) of the end-to-end signal-to-noise ratio (SNR). Interestingly, our results indicate that, except for the first user (nearest user), in a high-SNR regime the ABER achieves a performance floor that depends solely on the user's power allocation coefficient and on the type of modulation, and not on the channel statistics or the amount of transmit and receive antennas. To the best of the authors' knowledge, no performance analyses have been reported in the literature for the considered scenario. The validity of all our expressions is confirmed via Monte-Carlo simulations.
\end{abstract}


\section{Introduction}

The constant demand for massive connectivity scenarios in the sixth-generation (6G) network systems poses greater challenges than technologies such as orthogonal multiple access (OMA) can manage  \cite{6g-ran,tutorial-5g-noma}.
In view of this, researchers have proposed several techniques with the potential of meeting the requirements of advanced wireless networks, such as 6G.
Among them, we highlight the non-orthogonal multiple access (NOMA) technique since it (i) significantly improves  spectral and energy efficiency, (ii) provides ultra-reliable and low-latency communications (URLLC), and (iii) enables massive connectivity \cite{noma-5g,noma-fra,noma-low-latency,noma-fow-lowlat,massive-noma-1,massive-noma-2}.
NOMA can be classified into two schemes: code-domain and power-domain multiplexing. In this paper, we will focus on the latter, although the same framework can be applied to the former with no substantial adjustments.
The difference between NOMA and its predecessor, OMA, is that NOMA allocates time-frequency resources that are superposed and transmitted to multiple users, where for each user a specific power coefficient is allocated based on their channel conditions. On the other hand, OMA allocates resources to each user either in time, in frequency, or in code \cite{Cheon17}.
In NOMA, each receiver can
decode its own information by using the method of successive interference cancellation (SIC)~\cite{nomavsoma,noma-lte-5g,noma-5g}.


To explore power consumption reduction as well as hardware complexity and, at the same time, to improve system capacity, several works analyzed the performance of a NOMA system over different antenna configurations, e.g., single-input single-output (SISO), multiple-input single-output (MISO), single-input multiple-output (SIMO), and multiple-input multiple-output (MIMO).
For instance, in \cite{siso-noma-kappa-mu-eta-mu}, the authors carried out an outage probability (OP) analysis for a SISO-NOMA system in a multi-user scenario operating over $\kappa$--$\mu$ and $\eta$--$\mu$ fading environments. 
In \cite{op-noma-kappa-mu-egc}, the authors derived an approximate closed-form solution for the OP of a SIMO-NOMA system considering equal-gain combining (EGC) receivers operating over $\kappa$--$\mu$ fading channels and employing the special case when $\kappa \to 0$ (i.e., the Nakagami-$m$ fading model).
The same authors proposed in \cite{simo-noma-eta-mu} an approximate analysis for the OP considering a SIMO-NOMA system and employing $\eta$--$\mu$ fading channels, selection combining (SC), maximal-ratio combining (MRC), and EGC receivers. 
In \cite{miso-noma-tas-5g} and \cite{secure-miso-noma-tas}, an analysis of performance and security for a MIMO-NOMA system using a transmit antenna selection (TAS) scheme and Rayleigh fading was presented.
In \cite{op-noma-mimo-tasmaj-mrc-19}, the authors calculated the OP for a MIMO-NOMA system by means of Monte-Carlo simulations and employing a majority-based TAS (TAS-maj) technique, MRC receivers, and Rayleigh fading. 
In \cite{Alqahtani21} and \cite{ElHalawany20}, the authors derived exact expressions for the OP, the average bit error rate (ABER), and the ergodic capacity (EC) of a SISO-NOMA system in a two-user scenario and considering $\kappa$--$\mu$ and shadowed $\kappa$--$\mu$ fading channels, respectively.
Recently, in \cite{noma-siso-miso-naka-tas}, the authors derived exact expressions for the OP, the ABER, and the EC for both a SISO-NOMA and MISO-NOMA (with TAS) system considering independent double Nakagami-$m$ fading channels.


According to what has been shown in the open literature, and to the best of our knowledge, no \emph{exact} performance analysis has been carried out employing NOMA, TAS, and EGC receivers for any of the well-known fading distributions, not even for Rayleigh.
This work is the first of its kind analyzing, in an exact manner, the performance of a NOMA system considering the two aforementioned diversity schemes: TAS and EGC receivers.
For the analysis, we consider the Weibull fading model as it can accurately describe the non-linearities of the propagation medium in addition to encompassing, as special cases, important fading scenarios such as Rayleigh and exponential models.

The contributions of this work are summarized as follows:
\begin{enumerate}
    \item Novel exact expressions for the probability density function (PDF) and cumulative distribution function (CDF) of the signal-to-noise ratio (SNR) of a MU-MIMO-NOMA system operating over independent and identically distributed (i.i.d.) Weibull fading channels.
    \item New exact formulations for key performance metrics, namely, OP and ABER.
    Two important remarks are in order. Due to the versatility of the Weibull fading model, the exact OP and the ABER of MRC receivers can also be found from our analytical findings by replacing $k$ by $k/2$, where $k$ is the shape parameter of the Weibull distribution.
    Moreover, if we consider a single user (a non-NOMA system) and a single transmitting antenna, then our expressions provide an exact analysis in terms of OP and ABER for a SIMO system over Weibull fading channels, which so far have only been found as approximate, limited, or computationally expensive solutions \cite{Brennan59,Beaulieu90,weibull-egc,Ismail06,weibull-mrc,weibull-tas-mrc,Zhang21}. 
    \item Asymptotic closed-form expressions for the OP and the ABER. We show that, for the OP, the diversity order equals $(k M N)/2$, where $M$ and $N$ are the number of transmit and receive antennas, respectively. In contrast, for the ABER, the diversity order remains $(k M N)/2$ only for the first user, while being nil for the remaining ones, with the corresponding performance floor at high SNR solely depending on the user's power allocation coefficient and the employed modulation.   
\end{enumerate}

The remainder of this manuscript is organized as follows. Section~\ref{sec:system-model} introduces the MU-MIMO-NOMA system model.
Section \ref{sec: SNR's Statistics} obtains novel exact expressions for the SNR' statistics.
Section~\ref{sec:performance-analysis} derives exact and asymptotic expressions for the OP and the ABER.
Section~\ref{sec:numerical-results} discusses the representative numerical results. Finally, Section~\ref{sec:conclusion} concludes this paper.

\section{System Model}
\label{sec:system-model}

A downlink MU-MIMO-NOMA system with a single transmitter node $T$ and multiple receiving users $R_l$ ($l \in \left\{1,2,\hdots,U \right\}$) is considered, as shown in Fig.~\ref{fig:sys-model}.
The transmitter and receiving nodes are equipped with $M$ and $N$ antennas, respectively.
\begin{figure}[t]
\centering
\includegraphics[scale=0.4]{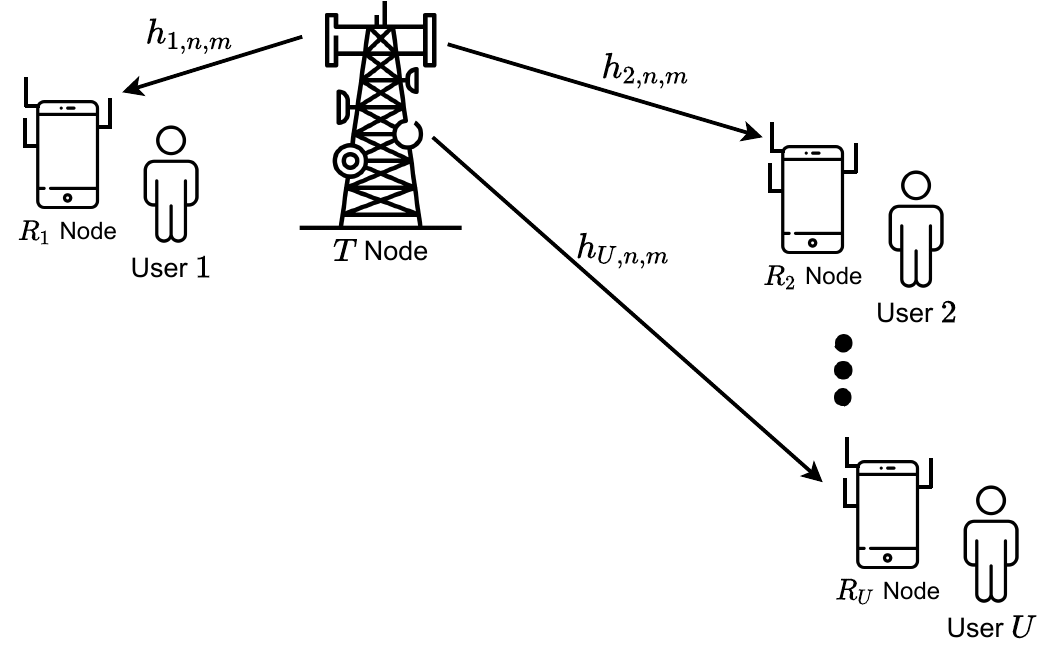}
\caption{Downlink MU-MIMO-NOMA system model.}
\label{fig:sys-model}
\end{figure}
Based on NOMA principle, $T$ simultaneously serves the multiple users over the same time and frequency resources.
Users with poor channel conditions are allocated with high power coefficients, and vice-versa. 
The complex channel coefficient between the $m$-th transmitting antenna and the $n$-th receiving antenna of the $l$-th user is denoted by $h_{l,n,m}$.
Herein, we assume that the envelope of each channel coefficient (i.e., $|h_{l,n,m}|$) follows a Weibull distribution with shape and scale parameters $k$ and $\lambda$, respectively.
Also, we assume that all channels experience i.i.d. fading and that perfect SIC is implemented to decode the superimposed signals \cite{2014sic}.
Without loss of generality, we assume that $|h_{U,n,m}| \leq ... \leq |h_{l,n,m}| \leq ... \leq |h_{1,n,m}|$.\footnote{The decoding order of SIC is given by the channel ordering \cite{noma-power-allocation}.}
Accordingly, the users' power allocation coefficients can be sorted as $\beta_U \geq ... \geq \beta_l \geq ... \geq \beta_1$, obeying $\sum_{j=1}^{U} \beta_j = 1$.

The superimposed information signal sent by node $T$ is given by $s=\sum_{j=1}^{U} \sqrt{P_j}x_j$ where $x_j$ is the information sent to each user, $P_j=\beta_j P_s$ is the transmit power, and $P_s$ is the total transmit power.
To leverage the benefits of multiple antennas (i.e., to exploit antenna
diversity), we employ the TAS strategy at transmission and EGC receivers at reception. 
Thus, the received signal at the $l$-th user can be written as
\begin{align}
    \label{eq: baseband signal}
    y_{l,n,m^*}=h_{l,n,m^*}\;\sum_{j=1}^{U} \sqrt{\beta_j P_s}x_j +n_{l,n},
\end{align}
where $n_{l,n}$ is the complex additive white Gaussian noise (AWGN) with zero mean and variance $\sigma_0^2$ present at the $n$-th antenna of the $l$-th user, and $m^*$ denotes the single antenna of node T selected for transmission according to the following criterion \cite{tas-ref}:
\small
\begin{align}
\label{eq:arg-max}
    m^*= \text{arg}\;\; \underset{1\leq m\leq M}{\text{max}} \;\;\left( \sum_{n=1}^{N} |h_{u,n,m}|  \right)^2.
\end{align}
\normalsize

\section{SNR's Statistics}
\label{sec: SNR's Statistics}


According to \eqref{eq: baseband signal}, the instantaneous SNRs at the first (nearest user) and $u$-th ($u \in \left\{2,3,\hdots,U \right\}$) users are respectively given by
\small
\begin{subequations}
\begin{align}
\label{eq:snr-u1}
    \chi_1=&\frac{\beta _1 \rho  \Psi _1^2}{N}\\
    \chi_{u}=&\frac{\beta_u\,\rho\,\Psi_u^2}{N + \rho\,\Psi_u^2 \vartheta_u}, \label{eq:snr}
\end{align}
\end{subequations}
\normalsize
where $\rho=P_s/\sigma_0^2$ denotes the transmit SNR, $\vartheta_u=\sum _{j=1}^{u-1} \beta_j$, $\Psi_1=\sum_{n=1}^{N} |h_{1,n,m^*}|$, and $\Psi_u=\sum_{n=1}^{N} |h_{u,n,m^*}|$ with $|h_{u,n,m^*}|$ being the independent and identically distributed (i.i.d.) Weibull fading envelopes satisfying the criterion in \eqref{eq:arg-max}.

Taking into account that all channels undergo independent fading, the CDF of $\Psi_u$ can be found as
\small
\begin{align}
    \label{eq: Probability definition}
    F_{\Psi_u}(\psi_u)= \left( \text{Pr} \left[ \sum_{n=1}^{N} |h_{u,n,m}| \leq \psi_u \right] \right)^M,
\end{align}
\normalsize
where $\text{Pr} \left[ \cdot \right]$ denotes probability.

Capitalizing on \cite[Proposition 1]{sum-weibull-21}, we can rewrite \eqref{eq: Probability definition} as
\small
\begin{align}
    \label{eq: sum-max-simple}
    F_{\Psi_u}(\psi_u)= \left(\frac{k}{\lambda^k }\right)^{N M} \left( \sum _{i=0}^{\infty } \frac{\delta_i \,\psi_u^{i k+ k N}}{\Gamma (i k+N k+1)} \right)^M,
\end{align}
\normalsize
where $\Gamma(\cdot)$ is the gamma function~\cite[eq. (6.1.1)]{abramowitz72}, and the coefficients $\delta_i$ can be obtained recursively as
\small
\begin{subequations}
\label{eq: delta i}
\begin{align}
    \delta _0=&\,\Gamma (k)^N\\
    \delta _i=&\,\frac{1}{i\Gamma (k)} \sum _{p=1}^i \frac{\delta _{i-p} (-i+p N+p) \Gamma (p k+k) \left(-\left(\frac{1}{\lambda }\right)^k\right)^p }{p!}.
\end{align}
\end{subequations}
\normalsize


\noindent
From \eqref{eq: sum-max-simple}, we let
\small
\begin{align}
    \label{eq: Differentiation}
    \left( \sum _{i=0}^{\infty}   \psi_u^{k i} \eta_i \right)^M =\sum _{i=0}^{\infty} \psi_u^{k i} \xi_i,
\end{align}
\normalsize
in which $\eta_i =\delta_i/\Gamma (i k+N k+1)$.

Now, we make use of the following differential equation:
\small
\begin{align}
    \label{eq: differential equation}
    \varrho \left(\varrho^{M }\right) '=M \varrho^{M} \varrho',
\end{align}
\normalsize
where $\varrho=\sum _{i=0}^{\infty} \psi_u^{k i} \eta_i$, $\varrho^{M} = \sum _{i=0}^{\infty}  \psi_u^{k i} \xi_i$ and the ``apostrophe" denotes derivative with respect to $\psi_u^{k}$.

After solving \eqref{eq: differential equation}, the coefficients $\xi_i$ can be calculated as
\small
\begin{subequations}
\label{eq: xi coefficients}
\begin{align}
    \xi _0=&\left(\frac{\delta _0}{\Gamma (k N+1)}\right)^M\\
    \xi _i=&\frac{\Gamma (k N+1)}{i\,\delta_0} \sum _{q=1}^i \frac{(-i+q M+q)\,\delta_q\,\xi_{i-q}}{\Gamma (q k+N k+1)},\ \ \ \  i\geq 1.
\end{align}
\end{subequations}
\normalsize

Finally, replacing \eqref{eq: Differentiation} and \eqref{eq: xi coefficients} into \eqref{eq: sum-max-simple}, we can express the CDF of $\Psi_u$ as follows
\small
\begin{align}
\label{eq: cdf-psi}
    F_{\Psi_u}(\psi_u)=& \left(\frac{k}{\lambda^k }\right)^{N M} \sum _{i=0}^{\infty } \xi_i\,\psi_u^{k (i + N M)}.
\end{align}
\normalsize
Then, by taking the derivative of \eqref{eq: cdf-psi} with respect to $\psi_u$, one attains the PDF of $\Psi_u$, i.e.,
\small
\begin{align}
    \label{eq: pdf-psi}
    f_{\Psi_u}(\psi_u)=& \,\frac{k^{N M+1}}{\psi_u}\left(\frac{\psi_u}{\lambda }\right)^{k N M} \sum _{i=0}^{\infty } \xi_i (i+N M) \psi_u^{i k}.
\end{align}
\normalsize



Using \eqref{eq:snr}, the CDF of $\chi_u$ can be found  as
\small
\begin{align}
    F_{\chi_u}(\chi_u)=&\,\text{Pr}\left[\Psi_u \leq \sqrt{\frac{N \chi_u}{\rho  \left(\beta_u-\chi_u \vartheta_u \right)}}\,\right]\nonumber\\
    =&\,F_{\Psi_u}\left( \sqrt{\frac{N \chi _u}{\rho  \left(\beta_u-\chi_u \vartheta_u\right)}}\,\right).
\label{eq:prev-cdf-snr}
\end{align}
\normalsize

Then, by employing \eqref{eq: cdf-psi} and \eqref{eq:prev-cdf-snr}, the CDF of $\chi_u$ can be finally expressed as
\small
\begin{align}
\label{eq:cdf-snr}
    F_{\chi_u}(\chi_u)=&\;\left(\frac{k}{\lambda^k}\right)^{N M}
    \sum _{i=0}^{\infty } \xi _i \left(\frac{N \chi_u}{\rho (\beta _u-\chi_u \vartheta_u)}\right)^{\frac{k}{2}(i+N M)}.
\end{align}
\normalsize

After differentiating \eqref{eq:cdf-snr} with respect to $\chi_u$, the PDF of $\chi_u$ can be found as
\normalsize
\small
\begin{align}
\label{eq:pdf-snr}
    f_{\chi_u}(\chi_u)&=\,\frac{\beta_u \, k^{1+N M}}{2\,\chi_u \left(\beta_u-\chi_u \vartheta_u\right)} \left(\frac{1}{\lambda }\right)^{k N M}\nonumber\\
    &\times \sum _{i=0}^{\infty } \xi_i\, (i+M N) \left(\frac{ N \chi_u}{\rho (\beta_u-\chi_u \vartheta_u)}\right)^{\frac{k}{2}(i+N M)}.
\end{align}
\normalsize

Following a similar approach as in \eqref{eq:cdf-snr} and \eqref{eq:pdf-snr}, the CDF and PDF of the SNR at the first user can be respectively obtained as\footnote{As the series in \eqref{eq: sum-max-simple} convergences absolutely \cite[Appendix]{sum-weibull-21}, then any further manipulation or transformation over \eqref{eq: sum-max-simple} will result in another absolute convergence series.
The rest of the derivations follow from here.}
\small
\begin{subequations}
\begin{align}
\label{eq:cdf-snr-u1}
    F_{\chi_1}(\chi_1)=&\left(\frac{k}{\lambda ^k}\right)^{M N} \sum _{i=0}^{\infty } \xi _i \left(\frac{N \chi _1}{\beta _1 \rho }\right)^{\frac{k}{2}(i+M N)}\\
    f_{\chi_1}(\chi_1)=&\frac{k^{M N+1}}{2\,\chi_1\,\lambda ^{k M N}} \sum _{i=0}^{\infty } \xi_i (i+M N) \left(\frac{N \chi _1}{\beta _1 \rho }\right)^{\frac{k}{2} (i+M N)}. \label{eq:pdf-snr-u1}
\end{align}
\end{subequations}
\normalsize

\section{Performance Analysis}
\label{sec:performance-analysis}
In this section, we use our derived formulations to analyze the performance of an EGC receiver subject to Weibull fading.

\subsection{Outage Probability}
The OPs for the first and $u$-th users are respectively defined as the probability that $\chi_1$ and $\chi_u$ fall below a specified threshold, i.e.,
\small
\begin{align}
\label{eq:def-pout}
    P_{\text{out},\nu}\overset{\Delta}{=}& \,\text{Pr}[\chi_\nu \leq \gamma_\nu]=  F_{\chi_\nu}(\gamma_\nu),
\end{align}
\normalsize
where $\nu \in \left\{1, u \right\}$ denotes the associated user (recall that $u \in \left\{2,3,\hdots,U \right\}$).

Now, from \eqref{eq:cdf-snr} and \eqref{eq:def-pout}, the OPs for the first and $u$-th users are respectively given by 
\small
\begin{align}
    \label{eq:pout_1-series}
   P_{\text{out},1}=&\left(\frac{k}{\lambda ^k}\right)^{N M} \sum _{i=0}^{\infty } \xi _i \left(\frac{N \gamma_1}{\rho \beta _1}\right)^{\frac{k}{2}(i+N M)}  \\ \label{eq:pout-series}
    P_{\text{out},u}=&\;\left(\frac{k}{\lambda^k}\right)^{N M}  \sum _{i=0}^{\infty } \xi _i \left(\frac{ N \gamma_u}{\rho(\beta_u-\gamma_u\, \vartheta_u)}\right)^{\frac{k}{2}(i+N M)}.
\end{align}
\normalsize

Moreover, we analyze the system performance in a high SNR regime (i.e., when $\rho \to \infty$). Then, since the first term dominates the series in \eqref{eq:pout_1-series} and \eqref{eq:pout-series} (i.e., the term associated with $i=0$), the asymptotic OPs for the first and $u$-th users can be expressed as
\small
\begin{align}
\label{eq:pout-asymp}
    P_{\text{out},\nu}\simeq  \left(O_{\textmd{c},\nu}\,\rho  \right)^{-O_{\textmd{d},\nu}},
\end{align}
\normalsize
where $\simeq$ denotes ``asymptotically equal to''; $O_{\textmd{d},1}=k M N/2$ and $O_{\textmd{d},u}= k M N/2 $ are the diversity gains for the first and $u$-th users, respectively; and
\small
\begin{align}
\label{eq:pout-gains_1}
    O_{\textmd{c},1}&= \frac{\lambda ^2  \beta_1}{ N \gamma_1} \left( \frac{\Gamma (k+1)^{ N}}{\Gamma (k N+1)}\right)^{-\frac{2}{k N}} \\ \label{eq:pout-gains}
    O_{\textmd{c},u}&= \frac{\lambda ^2  \left(\beta_u-\gamma_u \vartheta_u\right)}{ N \gamma_u} \left( \frac{\Gamma (k+1)^{ N}}{\Gamma (k N+1)}\right)^{-\frac{2}{k N}}
\end{align}
\normalsize
are the coding gains for the first and $u$-th users, respectively.




\subsection{ABER}
The ABERs for a pre-detection EGC receiver for the first and $u$-th users are respectively given by \cite[eq. (9.61)]{simon2004}
\small
\begin{align}
\label{eq:aber user1-def}
    P_{b,1}=&\frac{1}{2} \int_0^{\infty } \text{erfc}\left(\sqrt{\frac{\mathcal{A}\,  \beta _1 \, \rho \,   \psi _1^2}{N}}\right) \mathit{f}_{\Psi_1}(\psi_1) \, \text{d}\psi_1 \\ \label{eq:aber-def}
    P_{b,u}=&\frac{1}{2} \int_0^{\infty } \text{erfc}\left(\sqrt{\frac{\mathcal{A}\, \beta_u\,\rho\,\psi_u^2}{N+\rho\,\psi_u^2\,\vartheta_u}}\right) \mathit{f}_{\Psi_u}(\psi_u) \, \text{d}\psi_u,
\end{align}
\normalsize
in which $\text{erfc}(\cdot)^\text{3}$ is the complementary error function \cite[eq. (7.1.2)]{abramowitz72}, and $\mathcal{A}$ is a modulation-dependent parameter.
Considering the $u$-th user, the  improper integral in \eqref{eq:aber-def} can be expressed in terms of the limit when $\tau$ approaches infinity, i.e.,
\small
\begin{align}
    \label{eq: limit-form}
    P_{b,u}=\frac{1}{2} \lim_{\tau \to \infty }  \int_0^{\tau} \text{erfc}\left(\sqrt{\frac{\mathcal{A}\, \beta_u\,\rho\,\psi_u^2}{N+\rho\,\psi_u^2\,\vartheta_u}}\right) \mathit{f}_{\Psi_u}(\psi_u ) \, \text{d}\psi_u.
\end{align}
\normalsize

Replacing \eqref{eq: pdf-psi} in \eqref{eq: limit-form} and then changing the order of integration, we get
\small
\begin{align}
    \label{eq: ex_steps}
    P_{b,u}=&\frac{k^{N M+1}}{2}\left(\frac{1}{\lambda }\right)^{k N M}\sum _{i=0}^{\infty } \xi_i (i+N M)\nonumber\\ 
    \times&\lim_{\tau \to \infty}  \int_0^{\tau} \text{erfc}\left(\sqrt{\frac{\mathcal{A}\, \beta_u\,\rho\,\psi_u^2}{N+\rho\,\psi_u^2\,\vartheta_u}}\right) \psi_u^{k (i+N M)-1} \, \text{d}\psi_u.
\end{align}
\normalsize

Finally, integrating by parts and after some algebraic manipulations with the aid of ~\cite[eq. (4.2.1)]{abramowitz72}, the ABER for the $u$-th user can be obtained as
\small
\begin{align}
\label{eq:aber-series}
    P_{b,u}=&\frac{k^{M N}}{2\,\lambda^{k M N}}\nonumber\\
    &\sum _{i=0}^{\infty }   \xi_i \lim_{\tau \to \infty } \left[ \tau^{k (i+M N)} \left(\text{erfc}\left(\varsigma(\tau)\right)+\frac{2\,\zeta(i,\tau)}{\sqrt{\pi }}\right) \right],
\end{align}
\normalsize
where $\varsigma(\tau)$ and $\zeta(i,\tau)$ are auxiliary functions given by
\small
\begin{align}
    \varsigma(\tau)=&\sqrt{\frac{\mathcal{A}\,\rho \,\tau^2\,\beta_u}{N+\rho \,\tau^2 \vartheta_u}}\\
    \zeta(i,\tau)=&\sum _{j=0}^{\infty } \frac{\tau ^{2 j+1}}{j!\,(\varepsilon(i,j)+1)} \left(-\sqrt{\frac{\mathcal{A} \rho \beta_u}{N}}\right)^j \nonumber \\
    &\times\, _2F_1\left(j+\frac{3}{2},\frac{\varepsilon(i,j)+1}{2};\frac{\varepsilon(i,j)+3}{2};-\frac{\tau ^2 \vartheta_u}{N\, \rho^{-1}}\right),
\end{align}
\normalsize
in which $\varepsilon(i,j)=2 j+i k+k M N$ and $\, _2F_1(\cdot,\cdot; \cdot;\cdot)$\footnote{It is worth noting that $\text{erfc}(\cdot)$ and $\, _2F_1(\cdot,\cdot; \cdot;\cdot)$ can be
quickly and efficiently evaluated in any mathematical software.} is the Gauss hypergeometric function~\cite[Eq. (15.1.1)]{Olver10}. 

To ease the numerical calculation, \eqref{eq:aber-series} can be accurately approximated as
\small
\begin{align}
    \label{eq: aber-haprox}
    P_{b,u} \approx &\frac{k^{M N}}{2\,\lambda^{k M N}} \sum _{i=0}^{\infty }   \xi_i  \left(\mathit{a}^{\dagger} \right)^{k (i+M N)} \left(\text{erfc}\left(\varsigma_{(\mathit{a}^{\dagger})}\right)+\frac{2\,\zeta_{(i,\mathit{a}^{\dagger})}}{\sqrt{\pi }}\right),
\end{align}
\normalsize
where $\mathit{a}^{\dagger}$ is an accuracy-dependent parameter.
The higher the values of $\mathit{a}^{\dagger}$, the higher the accuracy. 
A value of $\mathit{a}^{\dagger}=35$  guarantees a relative error of less than $10^{-6}$, as will be seen in Section \ref{sec:numerical-results}.

Following the same derivation steps as in \eqref{eq:aber-series}, the exact ABER for the first user can be obtained as
\small
\begin{align}
\label{eq:aber-series-u1}
     P_{b,1}=\,\frac{k^{M N}}{2 \sqrt{\pi}\,\lambda ^{k M N}} \sum _{i=0}^{\infty } \frac{\xi_i\,\Gamma \left(\frac{i k+k M N+1}{2}\right)}{\left(\frac{\mathcal{A}\,\beta_1 \rho}{N}\right) ^{\frac{k}{2}(i+M N)}}.
\end{align}
\normalsize

An asymptotic ABER for the first user can also be found by using the first term in \eqref{eq:aber-series-u1}, resulting in
\small
\begin{align}
\label{eq:aber-asymp-u1}
    P_{\text{b},1}\simeq  \left(G_{\textmd{c},1}\,\rho  \right)^{-G_{\textmd{d},1}},
\end{align}
\normalsize
where $G_{\textmd{d},1}=k M N /2$ is the diversity gain and
\small
\begin{align}
\label{eq:code-gains-aber}
    G_{c,1}=\,\frac{\mathcal{A}\,\beta _1 \lambda ^2}{N} \left[\left(\frac{\Gamma \left(\frac{k M N+1}{2}\right)}{2 \sqrt{\pi }}\right)^M \frac{\Gamma (k+1)^N}{\Gamma (k N+1)}\right]^{-\frac{2}{k N}}
\end{align}
\normalsize
is the coding gain.
It is important to highlight that all exact expressions derived herein converge rapidly (i.e., with few summation terms) and are new in the literature.

\subsubsection{Minimum Achievable ABER for the $u$-th user}
Using a partial fraction decomposition into \eqref{eq:snr}, we obtain
\begin{align}
    \label{eq: SNR partial}
    \chi_{u}= \frac{\mathcal{A} \beta _u}{\vartheta _u}-\frac{\mathcal{A} N \beta _u}{\vartheta _u \left(N+\rho  \Psi ^2 \vartheta _u\right)}.
\end{align}
Noticing that in the high SNR regime (i.e., when $\rho \to \infty$), the second term vanishes.  Then, an asymptotic expression for the SNR can be obtained as
\small
\begin{align}
    \label{eq: ABER SNR ASYMP}
    \chi_{u} \simeq \frac{\mathcal{A} \beta _u}{\vartheta _u}.
\end{align}
\normalsize

Finally, substituting \eqref{eq: ABER SNR ASYMP} into \eqref{eq:aber-def}, a minimum achievable ABER (i.e., a performance floor at high SNR) for the $u$-th user can be found as
\small
\begin{align}
\label{eq:aber lower bound}
    P_{b,u}^{\text{min}}=\frac{1}{2}  \text{erfc}\left(\sqrt{\frac{\mathcal{A} \beta _u}{\vartheta _u}}\right).
\end{align}
\normalsize
Notice that the minimum achievable ABER for the $u$-th user is independent of the channel statistics (e.g., the type of fading) and the number of antennas in the diversity schemes (e.g., EGC, MRC, and TAS). Indeed, it only depends on the user's power allocation coefficient and the type of modulation.

\begin{figure}[t]
\centering
\includegraphics[scale=0.40]{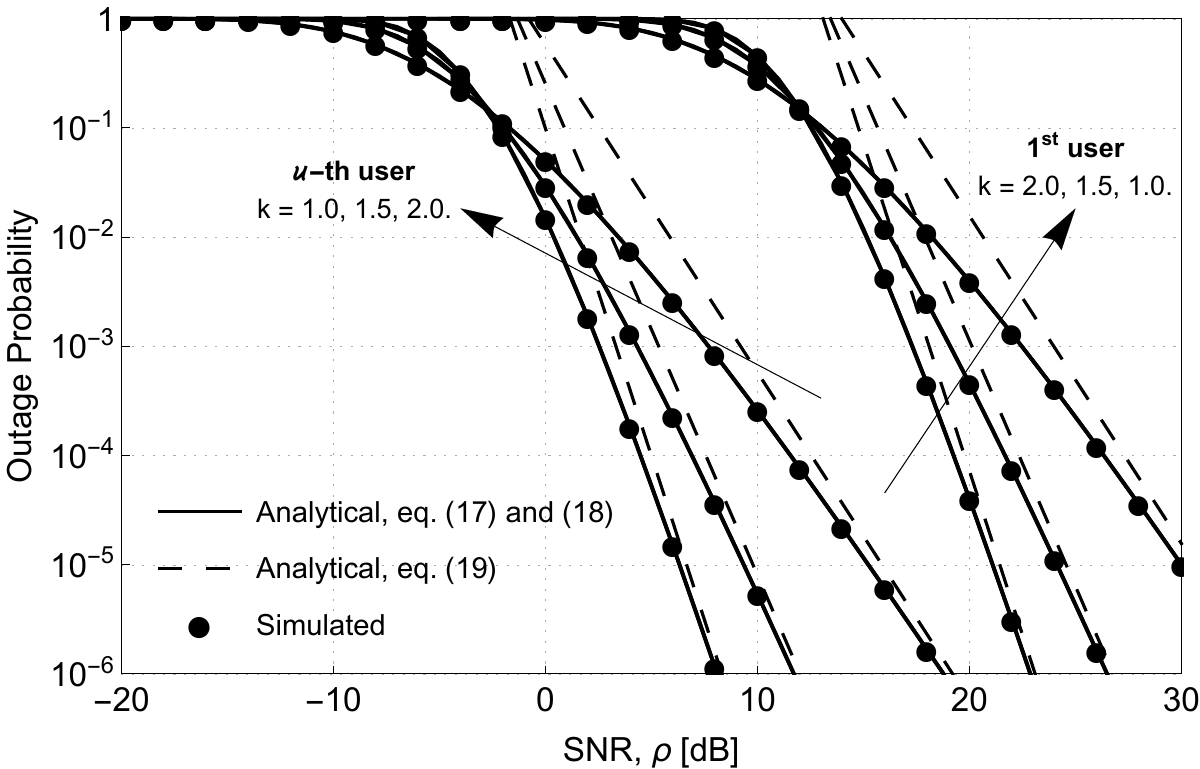}
\caption{OP versus SNR using $\lambda=2$, $N=2$, $M=3$, $\gamma_u=0$, $\beta_u=0.65$, $\beta_1=0.01$, $\vartheta_u=0.35$, and various values of $k$.}
\label{fig:pout-k}
\end{figure}
\begin{figure}[t]
\centering
\includegraphics[scale=0.4]{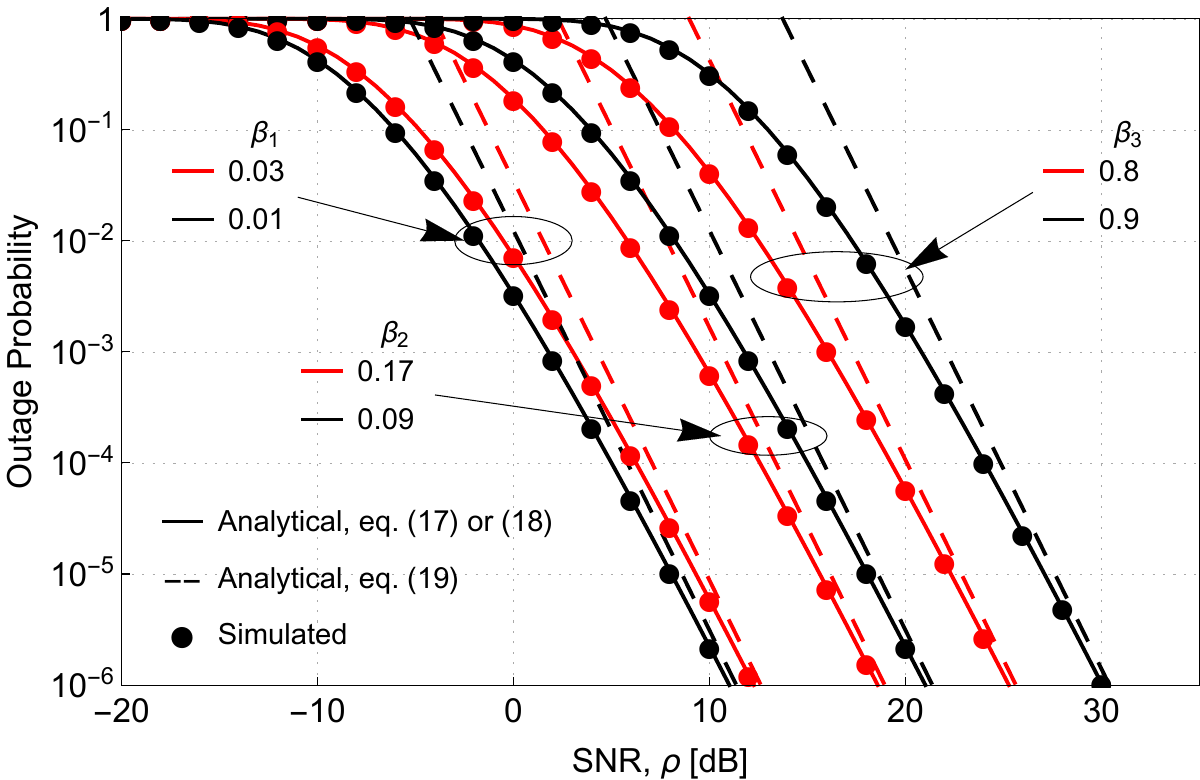}
\caption{OP versus SNR using $k=1.2$, $\lambda=2$, $N=2$, $M=3$, $\gamma_u=0$, and various values of $\beta_u$. This corresponds to two scenarios having three users each with power allocation coefficients $\beta_1$, $\beta_2$, and $\beta_3$.}
\label{fig:pout-bs-ps}
\end{figure}
\begin{figure}[t]
\centering
 \includegraphics[scale=0.4]{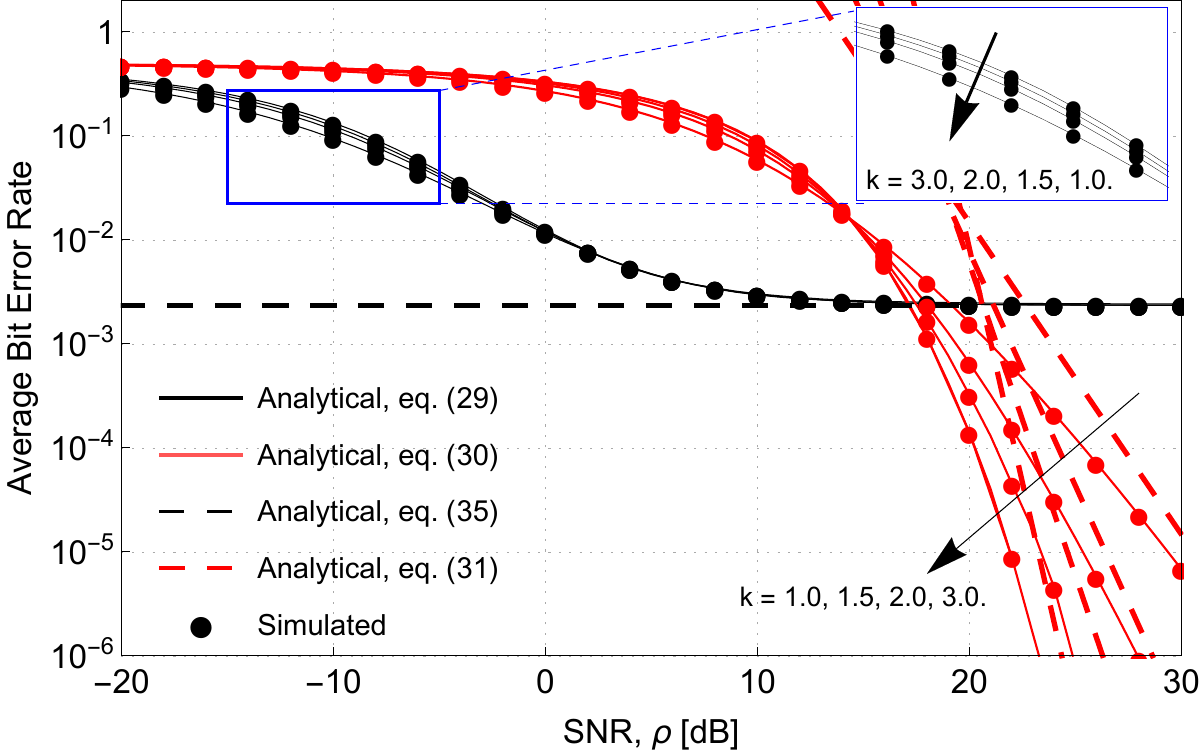}
 \caption{ABER versus SNR using $\lambda=2$, $N=2$, $M=3$, $\mathcal{A}=1$, $\beta_u=0.8$, $\vartheta_u=0.2$, $\beta_1=0.01$ and various values of $k$.}
\label{fig:aber-k}
\end{figure}
\begin{figure}[t]
\centering
\includegraphics[scale=0.4]{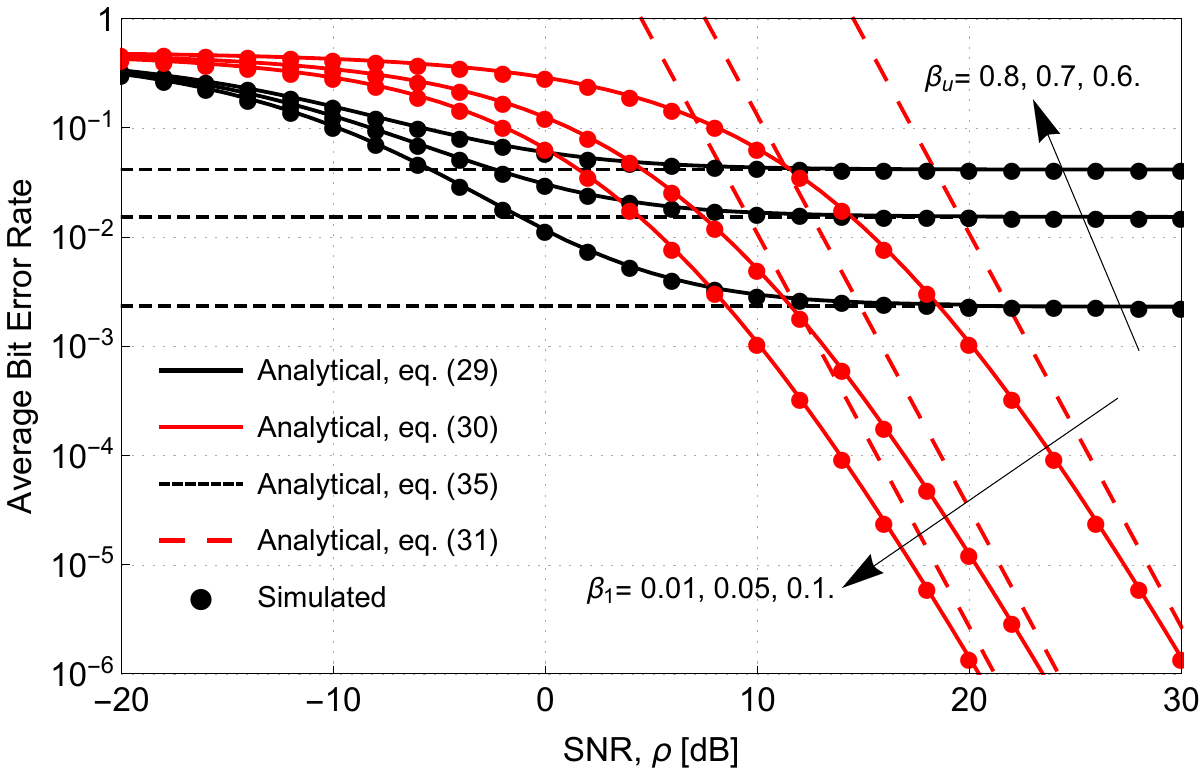}
\caption{ABER versus SNR using $k=1.2$, $\lambda=2$, $N=2$, $M=3$, $\mathcal{A}=1$ $\vartheta_u=1-\beta_u$ and various values of $\beta_u$ and $\beta_1$.}
\label{fig:aber-bs-ps}
\end{figure}
\section{Numerical Results}
\label{sec:numerical-results}

In this section, we corroborate our analytical findings through Monte-Carlo simulations.\footnote{The number of Monte-Carlo realizations was set to $10^7$. Also, we used a maximum of 200 terms in our derived series.}


Fig.~\ref{fig:pout-k} shows the OP in terms of the SNR for various values of shape parameter $k$. The figure indicates that for the first and $u$-th user, the system performance improves as $k$ increases and deteriorates otherwise.
The number of transmitting and receiving antennas, $N$ and $M$, also improves the system's reliability. However, the figures varying the number of antennas were omitted here due to space limitations.
The beneficial effects of increasing $k$, $N$, or $M$ is because the system's diversity order is equal to $(k N M)/2$. 

Fig.~\ref{fig:pout-bs-ps} shows the OP in terms of the SNR for two scenarios (curves in red for the first scenario and curves in black for the second scenario) with three different values of  $\beta_u$ (i.e., three users). More precisely, Fig. \ref{fig:pout-bs-ps} highlights the impact of the power allocation factor $\beta$ in a MU-MIMO-NOMA system. Since the majority of the power is allocated to the most distant user, lower SNR values are needed to achieve a high performance in terms of OP, as expected. Then, it is straightforward to see that the second user is placed in middle ground in terms of performance and that the nearest user shows the worst performance due to the low power coefficient assigned. 
Here, the importance of correct power allocation in NOMA systems is evidenced.


Fig.~\ref{fig:aber-k} depicts the ABER for multiple values of $k$ for the $u$-th and first user. For the $u$-th the figure indicates that higher values of the shape parameter $k$ lead to higher values of ABER, which deteriorates the system performance. For the first user the opposite occurs, i.e., as $k$ increases better ABER values are reached. Also, notice that regardless of the values of $k$, at high SNR regime the ABER reaches the minimum value (performance floor), given in \eqref{eq:aber lower bound} (the same occurs for different values of the scale parameter $\lambda$, omitted here due to space constraints). That is, the minimum ABER is independent of the fading parameters.
Moreover, for the $u$-th user, the figure also shows that for $k>1$ the ABER curves almost overlapped. Therefore, in scenarios where the fading channel follows a Weibull distribution with shape parameter $k>1$, an analytical (or simulated) ABER analysis can be simplified (or approximated) by considering $k=2$, i.e., Rayleigh fading.

Fig.~\ref{fig:aber-bs-ps} shows the ABERs for different values of $\beta_1$ and $\beta_u$. 
In particular, for the first user, it can be noticed that as the SNR increases the ABER decreases. On the other hand, for the $u$-th user, it can be seen that at high SNR, the ABER reaches a performance floor that depends on the power coefficients. Notice how the system improves as the power coefficients increase, i.e., the higher the power coefficient, the smaller the ABER. Also, notice the influence of the power coefficient parameters in the ABER, where higher $\beta_1$ and $\beta_u$ values produce smaller ABER values.

Finally, notice the perfect agreement between Monte-Carlo simulations and our analytical results, thereby corroborating our findings. Moreover, notice how our asymptotic formulations provide excellent fits in the high-SNR regime.

\section{Conclusion}
\label{sec:conclusion}
This work analyzed the performance of a MU-MIMO-NOMA system operating over Weibull fading channels. 
Exact formulations for the SNR's statistics, OP, and ABER were provided. 
An asymptotic analysis was also carried out to show how the physical parameters roughly affect the system performance. 
Analytical and numerical results indicate that the system performance improves as $k$, $M$, or $N$ increases. Moreover, it was shown that the minimum achievable ABER for the $u$-th user is independent of the fading scenario and of the number of antennas in the diversity schemes. In fact, it only depends on the user's power allocation coefficient and the type of modulation.

\bibliographystyle{IEEEtran}
\bibliography{mybib}

\end{document}